\documentclass[prd,showpacs,floats,nofootinbib,preprint,floatfix]{revtex4}

\usepackage{amsmath,amssymb}
\usepackage{graphicx}
\usepackage{bm}

\begin{document}

\title{Heavy quark bound states above $T_c$}

\author{W.M. Alberico, A. Beraudo, A. De Pace and A. Molinari}, 
\affiliation{Dipartimento di Fisica Teorica dell'Universit\`a di Torino and \\ 
  Istituto Nazionale di Fisica Nucleare, Sezione di Torino, \\ 
  via P.Giuria 1, I-10125 Torino, Italy}

\begin{abstract}
A comprehensive parameterization of the colour singlet heavy quark free
energy above $T_c$ is given, using the lattice data in quenched ($N_f=0$) and
unquenched ($N_f=2$ and $N_f=3$) QCD. The corresponding (temperature dependent)
potentials thus obtained are then inserted into the Schr\"odinger equation for
the charmonium and the bottomonium in the deconfined phase of QCD. 
The solution of the equation provides an estimate of the melting temperature
and of the radii for the different $c\bar{c}$ and $b\bar{b}$ bound states.
\end{abstract}


\date{July 7, 2005}

\maketitle

\section{Introduction}

The anomalous suppression of the $J/\psi$ production in heavy ion collisions,
which has been experimentally observed \cite{Na50,Na50bis} in the depletion
of the dilepton multiplicity in the region of invariant mass corresponding to
the $J/\psi$ meson, was proposed long time ago as a possibly unambiguous signal
of the onset of deconfinement \cite{satz}. Indeed in Ref. \cite{satz} it is
argued that charmonium states can only be produced in the first instants after
the nucleus-nucleus collision, before the formation of a thermalized QGP. Then,
in their path through the deconfined medium, the original $c\bar{c}$ bound
states tend to melt, since the binding (colour) Coulomb potential is screened
by the large number of colour charges. This, in turn, produces an anomalous
(with respect to normal nuclear absorption) drop in the $J/\psi$ yields. 

In this picture it is implicitly assumed that, once the charmonium dissociates,
the heavy quarks hadronize by combining with light quarks only
(recombination leading to a secondary $J/\psi$ production is neglected). This
assumption is certainly justified at the SPS conditions, due to the very small
number of $c\bar{c}$ pairs produced per collision ($N_{c\bar{c}}\sim 0.2$ in a
central collision), but at RHIC ($N_{c\bar{c}}\sim 10$) and LHC
($N_{c\bar{c}}\sim 200$) energies it is no longer warranted \cite{tew}.

Moreover in a hadronic collisions only about $60\%$ of the observed $J/\psi$'s
are directly produced, the remaining stemming from the decays of excited
charmonium states (notably the $\chi_c$ and the $\psi'$). Since each $c\bar{c}$
bound state dissociates at a different temperature, a model of {\em sequential
  suppression} was developed, with the aim of reproducing the $J/\psi$
suppression pattern as a function of the energy density reached in the heavy
ion collision (the highest temperatures and energy densities being reached in
the most central collisions) \cite{satz2,diga1,diga2,kar,Kar97}. 
SPS experimental data for Pb-Pb collisions at different centralities seem
indeed to support the dissociation pattern predicted by this model 
\cite{Na50,Na50bis}.

Alternative mechanisms for the $J/\psi$ production, like the Statistical
Coalescence Model (SCM), have also been proposed \cite{scm}. In the SCM
one assumes again that all the $c\bar{c}$ pairs are produced in hard processes
at the initial stage of the collision. Any heavy quark bound state, if present,
is assumed to melt in the QGP phase and the number of $c\bar{c}$ pairs in the
fireball is considered fixed. At the chemical freeze-out open and hidden
charmed hadrons are then produced with multiplicity {\em ratios}\footnote{Due
  to the low rate of inelastic reactions full chemical equilibrium cannot be
  reached by charmed hadrons: their total multiplicity measured at SPS stays
  well above the thermal value.} fixed by their masses, according to the laws
of statistical mechanics. Hence, in such a scheme, the measured $J/\psi$
multiplicity is not related to the presence of charmonium bound states in the
plasma phase, but to the statistical hadronization of the initially produced
$c$($\bar{c}$) (anti-)quarks. This may lead, at LHC energies, to a completely
different picture characterized by an enhanced charmonium production even if
all the $c\bar{c}$ bound states dissociate during the plasma phase. 

In any case the hypothesis that all the primary produced $J/\psi$'s melt during
the QGP lifetime is hardly realized at SPS conditions. Hence models have been
developed \cite{gran1,gran2} attempting to account both for the initial state
production, eventually subject to in-medium dissociation, and for the thermal
production at the hadronization. 

Concerning the heavy quarkonia in the QGP phase, recent lattice
data \cite{dat,asa1,asa2} in quenched approximation (hence neglecting effects
arising from virtual processes involving dynamical fermions), which display
narrow peaks for the charmonium spectral functions in the pseudoscalar and
vector channels (even up to $T\sim 2T_c$), seem to point to the existence of
heavy quark bound states up to temperatures above $T_c$\footnote{Actually, in
  Ref.~\cite{asa1} the analysis has been done for $s\bar{s}$ mesons.}. 
Clearly these results, if confirmed, would entail striking experimental
consequences.

Actually, the meson spectral functions cannot be measured directly on the
lattice. From the numerical simulations one gets the current-current
correlation function along the (imaginary) temporal direction on a finite
number of points. Such a correlator corresponds to the convolution of the meson
spectral function with a thermal kernel. The spectral function can then be
obtained only indirectly. With this aim in Refs.~\cite{dat,asa1,asa2} a
procedure called Maximum Entropy Method (MEM) has been adopted. 
Clearly, an independent check of the results obtained with MEM appears
desirable. This indeed is our scope in the present paper. 

For this purpose, we first extract from lattice data a heavy quark potential
accounting for thermal effects and then we solve the Schr\"odinger equation for
the charmonium (and bottomonium). 
As shown in Refs. \cite{mc,Kac1,Kac2,Kac3,Kac4}, from the Polyakov loop
correlation function it is possible to extract the free energy (in the
different color channels) of a heavy quark-antiquark pair placed at a distance
$r$ in a thermal bath of gluons and light dynamical fermions. Once a good
parameterization of the color singlet free energy is obtained, the entropy and
internal energy contributions can be disentangled. Since the quarks acting
as static sources of the color field are considered infinitely heavy, the
internal energy coincides with the potential. The latter is then inserted into 
the Schr\"odinger equation, from which the binding energy of the different
stable states --- if there are any --- and their evolution with the temperature
are obtained. 

Indeed a clear distinction between the $Q\overline{Q}$ free and potential
energies is necessary in order to get a reliable estimate of the quarkonium 
dissociation temperature $T_d$ in the different spin-parity channels.

In Refs.~\cite{diga1,diga2,wong1,wong2}, where the colour singlet {\em free
  energy} was directly inserted into the Schr\"odinger equation, the
dissociation temperatures $T_d=1.10T_c$ \cite{diga2} and $T_d=0.99T_c$
\cite{wong1} were found for the $J/\psi$, all the other charmonium states
melting well below $T_c$.

On the other hand, in Ref.~\cite{wong3}, where a parameterization of the
lattice color singlet {\em potential} (in the quenched approximation) was used,
the temperature $T_d$ for the {\em spontaneous dissociation} of the $J/\psi$
was estimated to occur at about $2 T_c$, a value even larger than the one
obtained from the spectral analysis performed in Refs.~\cite{dat,asa1,asa2}. 
Also results for different charmonium and bottomonium states have been reported
in Ref.~\cite{wong3}. The case $N_f=2$ was addressed in Ref.~\cite{shury},
where the $J/\psi$ meson was found to be bound till $T\sim 2.7 T_c$.

Actually, even if the potential supports the existence of bound states, other
physical processes may lead to the dissociation of the quarkonium. First, if
the $Q\overline{Q}$ binding energy is lower than the temperature --- and
assuming that the quarkonia have reached the thermal equilibrium with the
plasma --- a certain fraction of their total number will be thermally excited
to resonant states according to a Bose-Einstein distribution: such a process is
referred to as {\em thermal dissociation} \cite{wong1,wong3}. 
Furthermore, the collisions with the gluons and the light quarks of the plasma
may lead to the {\em collisional dissociation} of the quarkonium. In this
connection, the reaction $g+J/\psi\rightarrow c+\overline{c}$ was studied in
detail in Ref.~\cite{wong3}. Hence, in spite of the presence of a bound state
solution of the Schr\"odinger equation till $T\sim 2T_c$, the $J/\psi$ turned
out to be really stable only up to temperatures lower than $2t_c$ \cite{wong3}.
Clearly the two processes above mentioned are not encoded in the Polyakov loop
correlation function, where the heavy quarks act as unthermalized static
sources of the color field, and have to be accounted for {\em a posteriori}.

Of course, if the process $g+J/\psi\rightarrow c+\overline{c}$ can lead to the
dissociation of the charmonium, the same reaction can also occur in the
opposite direction. Hence a consistent calculation of $J/\psi$ multiplicity
implies the solution of a kinetic rate equation integrated over the lifetime of
the QGP phase in which both processes (dissociation and recombination) enter
\cite{wong3,raf}. To carry out this detailed balance calculation the knowledge
of the $J/\psi$ binding energy and wave function in the thermal bath turns out
to be an important input. This is of relevance because, as mentioned above, the
usual assumption in considering the $J/\psi$ suppression as a signature of
deconfinement is that its production can occur only in the very initial stage
of the collision. Really, if at SPS the role played by recombination is
numerically negligible, this is no longer true at RHIC as pointed out in
Ref.~\cite{raf}.

In any case in this paper we limit ourselves to check the existence of bound
state solutions of the Schr\"odinger equation. A quantitative study of the
dissociation and recombination processes is left for future work.

Here we take advantage of all the available lattice data, obtained not only in
quenched QCD ($N_f=0$), but also including two and, more recently, three light 
flavors. We are then in a position to study also the flavor dependence of the
dissociation process, a perspective not yet achieved by the parallel studies of
the spectral functions, which are, as already mentioned, only available in
quenched QCD.

The present paper is organized as follows.
In Sec.~\ref{sec:para} we present a parameterization of the color singlet
free energy lattice data for the cases $N_f=0$ \cite{Kac1}, $N_f=2$ \cite{Kac4}
and $N_f=3$ \cite{Petr}, from which the heavy quark potential is obtained.
In Sec.~\ref{sec:qqbar} we solve numerically the associated Schr\"odinger
equation at different temperatures for the charmonium and bottomonium states,
thus determining their dissociation temperature.
Finally, in Sec.~\ref{sec:concl} we present our conclusions.

\section{Parameterization of the lattice data}
\label{sec:para}

\begin{figure}[t]
\begin{center}
\includegraphics[clip,width=\textwidth]{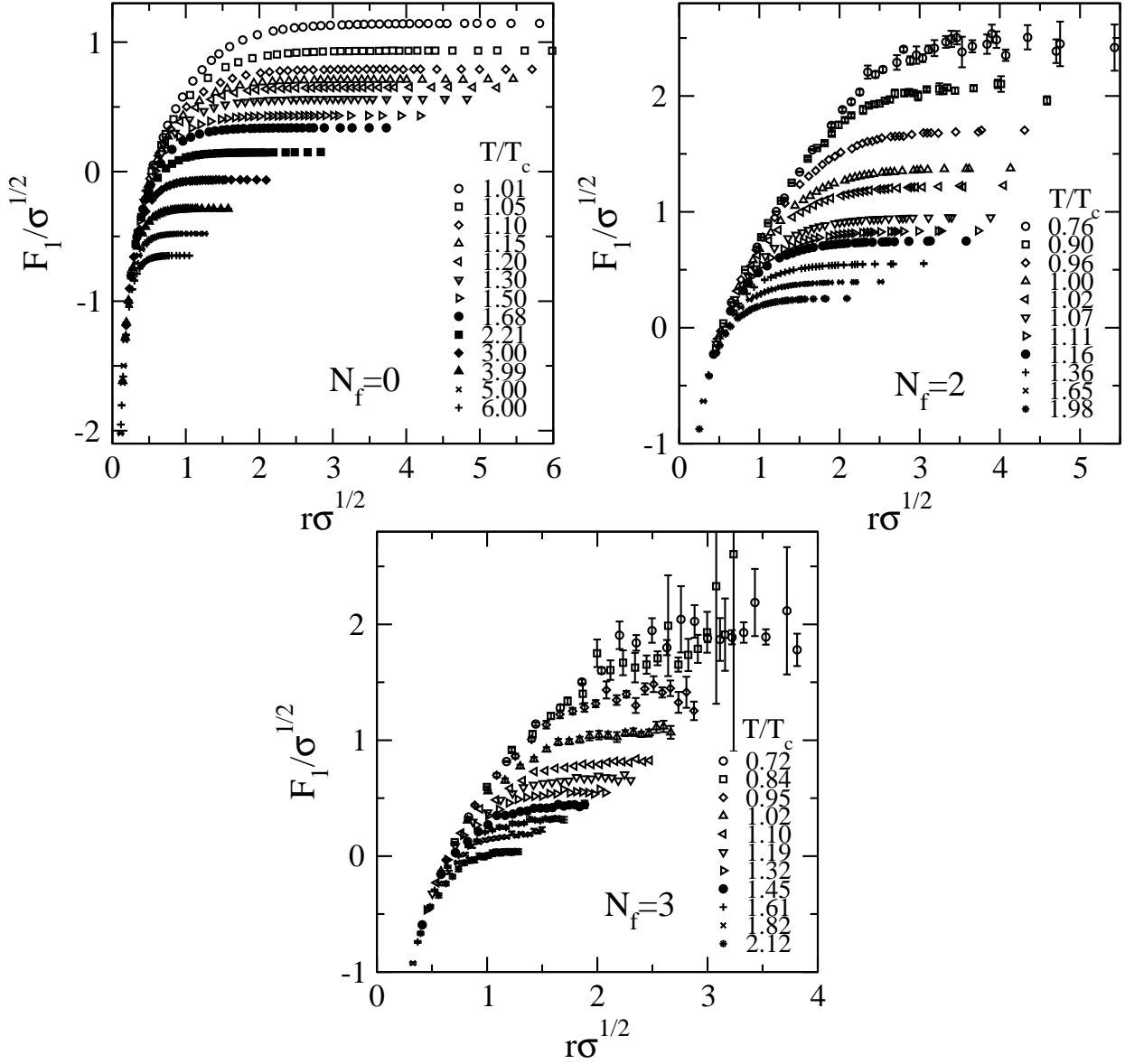}
\caption{A sample of the available data for $F_1(r,T)/\sqrt{\sigma}$ at
  different values of $T/T_c$ as a function of the separation $r\sqrt{\sigma}$
  of the $Q\bar{Q}$ sources. Data are taken from Ref.~\protect\cite{Kac1}
  ($N_f=0$), Ref.~\protect\cite{Kac4} ($N_f=2$) and Ref.~\protect\cite{Petr}
  ($N_f=3$). Dimensionless units are employed, with $\sqrt{\sigma}=420$~MeV
  ($N_f=0$ and $N_f=2$) and $\sqrt{\sigma}=460$~MeV ($N_f=3$). }
\label{fig:freeen}
\end{center}
\end{figure}

In this section we provide a unified parameterization of the lattice data for
the \emph{color singlet} $Q\bar{Q}$ \emph{free energy} $F_1$ in the case of
quenched \cite{Kac1}, 2-flavor \cite{Kac4} and 3-flavor \cite{Petr} QCD. The
lattice findings are shown (in dimensionless units) in Fig.~\ref{fig:freeen}.

For what concerns the critical temperature we assume the values $T_c=270$ MeV
($N_f=0$), $T_c=202$ MeV ($N_f=2$) and $T_c=193$ MeV ($N_f=3$) given in
Ref.~\cite{Kac4}.

The free energy on the lattice is defined up to an additive normalization
constant, which has to be fixed using some physical constraint.
In Refs.~\cite{Kac1,Kac4,Petr} it has been normalized to match, at the shortest
available distance for each temperature, the $T=0$ heavy quark potential. 
Such a normalization amounts to make the (reasonable) assumption that thermal
effects become negligible at very short distances.

In Refs.~\cite{Kac1,Kac4} the zero temperature heavy quark potential has been
determined through a best fit procedure of the available $T=0$ lattice data
with a Cornell-like parameterization\footnote{Note that the $1/r$ term accounts
  for two different physical processes: the perturbative one-gluon-exchange at
  short distances and the transverse string fluctuations at large distances.}:
\begin{equation}
  \frac{V(r)}{\sqrt{\sigma}}=-\frac{4}{3}\frac{\alpha}{r\sqrt{\sigma}}+
    \sqrt{\sigma}r,
\end{equation}
$\sigma$ representing the {\em string tension}. The values $\alpha=0.195(1)$
for the case $N_f=0$ and $\alpha=0.212(3)$ for the case $N_f=2$ are given in
Ref.~\cite{Kac4}, where the value $\sqrt{\sigma}=420$~MeV is employed to
translate the lattice results into physical units.

In Ref.~\cite{Petr} a similar parameterization (Cornell potential plus a
$1/r^2$ term to mimic the effects of asymptotic freedom at the shortest
distances reachable on the lattice) is employed for the case $N_f=3$. Note that
in that work the free energy is provided directly in physical units: however,
for the sake of comparison, we show it in dimensionless units, using for the
string tension the value $\sqrt{\sigma}=460$~MeV extracted from the
parameterization of the $T=0$ potential given in Ref.~\cite{Petr}.

In past calculations \cite{diga1,diga2,wong1,wong2} the free energy has been
often identified with the heavy quark potential and inserted directly into the
Schr\"odinger equation. 
However, a better candidate for a more appropriate finite temperature potential
is given by the internal energy of the $Q\bar{Q}$ system, defined by the well
known relation 
\begin{equation}
  F=U-TS,
\end{equation}
where
\begin{equation}
  U=-T^2\frac{\partial(F/T)}{\partial T}
\label{eq:pot}
\end{equation}
is the internal energy and 
\begin{equation}
  S=-\frac{\partial F}{\partial T}
\end{equation}
is the entropy.
One can see from the data in Fig.~\ref{fig:freeen} that the role played by 
the entropy is more relevant at large distances.

Getting the internal energy from the free energy involves a derivative of the
latter with respect to the temperature: it is thus clear that one needs an
accurate parameterization of the temperature dependence of $F$.

In order to establish a suitable form for this parameterization of the lattice
data, we first consider the two limits in which the underlying physics is
supposed to be known. 

At very short distances ($r\ll 1/T$) thermal effects are negligible and the
colour singlet free energy is dominated by the perturbative one-gluon exchange
with the typical behaviour:
\begin{equation}
  F_1(r,T)\underset{rT\ll1}{\sim}-\frac{4}{3}\frac{\alpha(r)}{r},
\label{eq:short}
\end{equation}
the coupling $\alpha$ depending only upon the $Q\bar{Q}$ separation.

On the other hand, for $T\gg T_c$, the large distances ($r\gg 1/T$) behaviour
of the free energy is expected to be dominated by the exchange of a resummed
electrostatic gluon leading to the expression \cite{nad}: 
\begin{equation}
  F_1(r,T)\underset{rT\gg1}{\sim}-\frac{4}{3}\frac{\alpha(T)}{r} 
    e^{-m_D(T)r}+F_1(r=\infty,T).
\label{eq:large}
\end{equation}
In this limit, the coupling $\alpha$ is a function of the temperature
and $m_D(T)$ is the Debye screening mass arising from the dressing of the
electrostatic gluon. 

In the two above limits, the running of the coupling is
determined by a Renormalization Group Equation (RGE), allowing to express
$\alpha=g^2/4\pi$ as a known (at least in the weak coupling regime) function of
an energy scale $\mu$. In the short distance limit the relevant energy scale is
given by the inverse of the distance ($\mu\sim1/r$), while for large
separations the major role in setting the scale is expected to be played by the
temperature ($\mu\sim T$).

Actually, in order to solve the Schr\"odinger equation for heavy quarkonia, one
really needs a parameterization of the free energy covering the whole range of
distances. For this purpose, on the basis of Eqs.~(\ref{eq:short}) and
(\ref{eq:large}) it appears convenient to cast the dependence of $F_1$ on $r$
and $T$ into the following functional form\footnote{Note that the mass $M(T)$
  appearing in this exponential will not necessarily coincide with the Debye
  screening mass of Eq.~(\protect\ref{eq:large}), the latter being determined 
  by fitting only the large distance data.}:
\begin{equation}
  F_1(r,T)=-\frac{4}{3}\frac{\alpha(r,T)}{r}e^{-M(T)r}+C(T). 
\end{equation}
In order to recover from the above the short and large distance limits given by
Eqs.~(\ref{eq:short}) and (\ref{eq:large}) one can, e.g., assume $\alpha$ to
depend on the following combination of $r$ and $T$: 
\begin{equation}
  \alpha(r,T)=\alpha(\mu=c_r/r+c_tT),
\label{eq:ansalpha}
\end{equation}
where $\alpha(\mu)$ is obtained by solving the RGE, while $c_r$ and $c_t$
are numerical coefficients to be fixed through a best fit of the data.

If supported by the data, Eq.~(\ref{eq:ansalpha}) would allow to interpolate
between the short distance regime ($r\ll 1/T$), where $\mu\sim 1/r$, and the
long distance one ($r\gg 1/T$), where on the contrary $\mu\sim T$.

\begin{figure}[t]
\begin{center}
\includegraphics[clip,width=\textwidth]{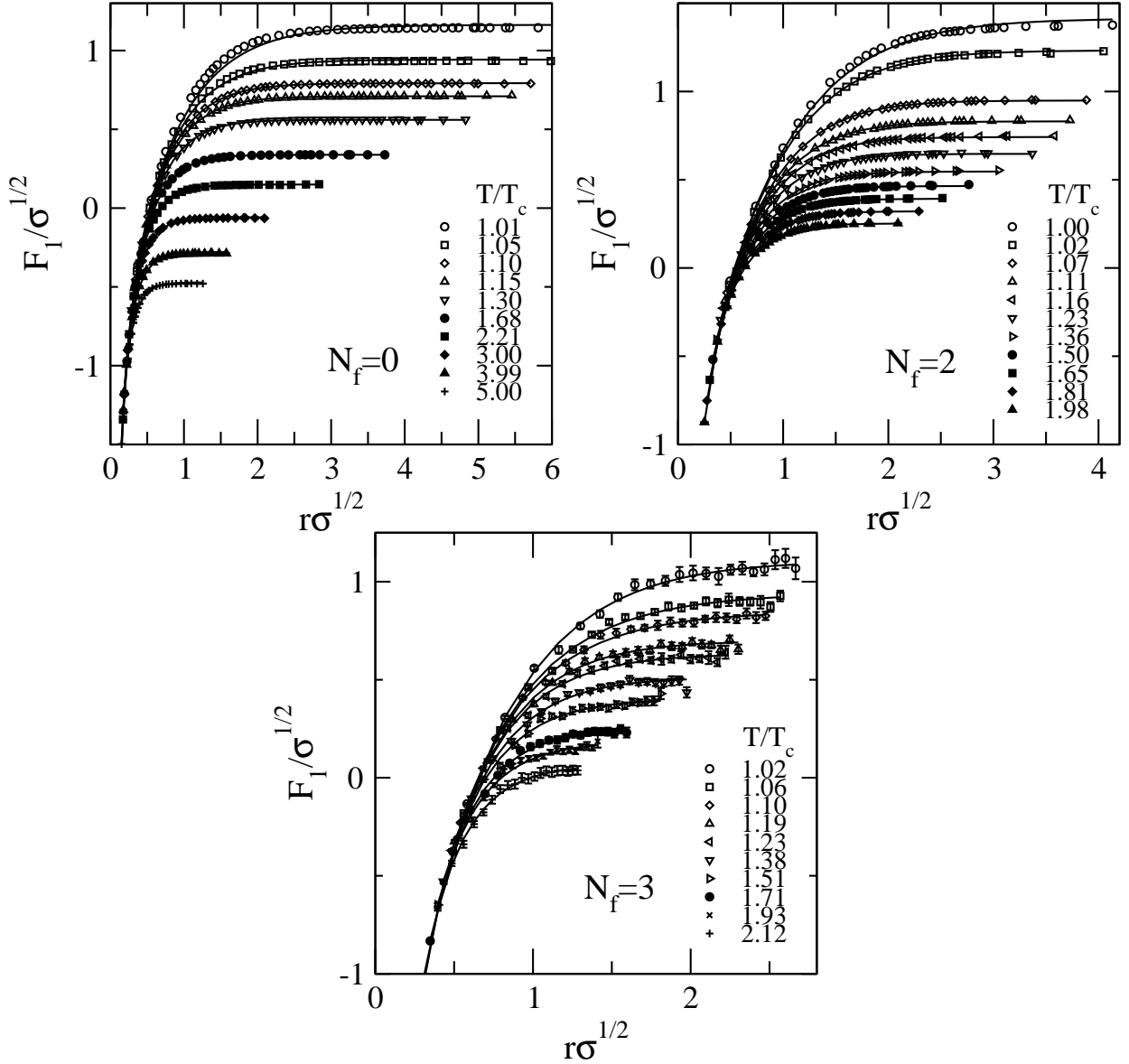}
\caption{The colour singlet free energy resulting from our fitting procedure 
  compared to the lattice data at different temperatures above $T_c$. }
\label{fig:fit3par} 
\end{center}
\end{figure}

We now describe in detail the fitting procedure we have employed. 
The data for the colour singlet free energy, taken from Ref.~\cite{Kac1}
($N_f=0$), Ref.~\cite{Kac4} ($N_f=2$) and Ref.~\cite{Petr} ($N_f=3$), are 
displayed in Fig.~\ref{fig:freeen} in dimensionless units, namely
$y=F_1/\sqrt{\sigma}$ and $x=r\sqrt{\sigma}$, for different values of the
temperature both below and above $T_c$. For each value of $T>T_c$ the data have
been parameterized with the following (four parameter) fitting function:
\begin{equation}
  y=-\frac{4}{3}\frac{\alpha(\tilde{\mu})}{x}e^{-a_3x}+a_0
    \qquad \textrm{with} \qquad \tilde{\mu}=\frac{a_1}{x}+a_2\,
\label{eq:fittingfunc}
\end{equation}
where for $\alpha(\tilde{\mu})$ we use the RGE result obtained with the 
two-loop QCD beta-function quoted in Appendix~\ref{app:short}, the
dimensionless variable $\tilde{\mu}$ being identified with the ratio
$\mu/\Lambda_{\textrm{QCD}}$. 

The fitting procedure yields a very mild dependence on $T$ for the parameters
$a_1$ and $a_2$. Actually, on the basis of the above discussion, one would have
expected the coefficient $a_2$ to scale linearly with the temperature, but our
finding might just signal that the range of temperatures spanned here is still
not in the asymptotic regime $T\gg T_c$. 
On the other hand, a constant value for $a_1$ is in agreement with the ansatz
of Eq.~(\ref{eq:ansalpha}). This has suggested the following procedure: 
a weighted average of the values obtained for $a_1$ has been performed,
yielding $a_1=0.2719(2)$ for $N_f=0$, $a_1=0.2687(7)$ for $N_f=2$ and
$a_1=0.2354(17)$ for $N_f=3$. 
We have then fitted again the data keeping $a_1$ fixed and using only $a_0$,
$a_2$ and $a_3$ as free parameters.  
This parameterization works well --- yielding values of $\chi^2$ per degree of
freedom of the order of 1 at all the temperatures --- and it is compared to the
data in Fig.~\ref{fig:fit3par}.

The finite temperature lattice data are limited to distances
$r\gtrsim0.1\div0.2$ fm. Since the data examined in this
work have been normalized assuming that at short distances thermal effects are
negligible, we should check that our parameterization does not introduce any
sizable (and spurious) temperature dependence for small values of $r$,
remaining in this region close to the $T=0$ perturbative potential (we remind
that at $T=0$ free energy and internal energy coincide). 
In this respect our choice of refitting the data keeping the coefficient $a_1$
fixed fulfills this requirement.

\begin{figure}[t]
\begin{center}
\includegraphics[clip,width=0.9\textwidth]{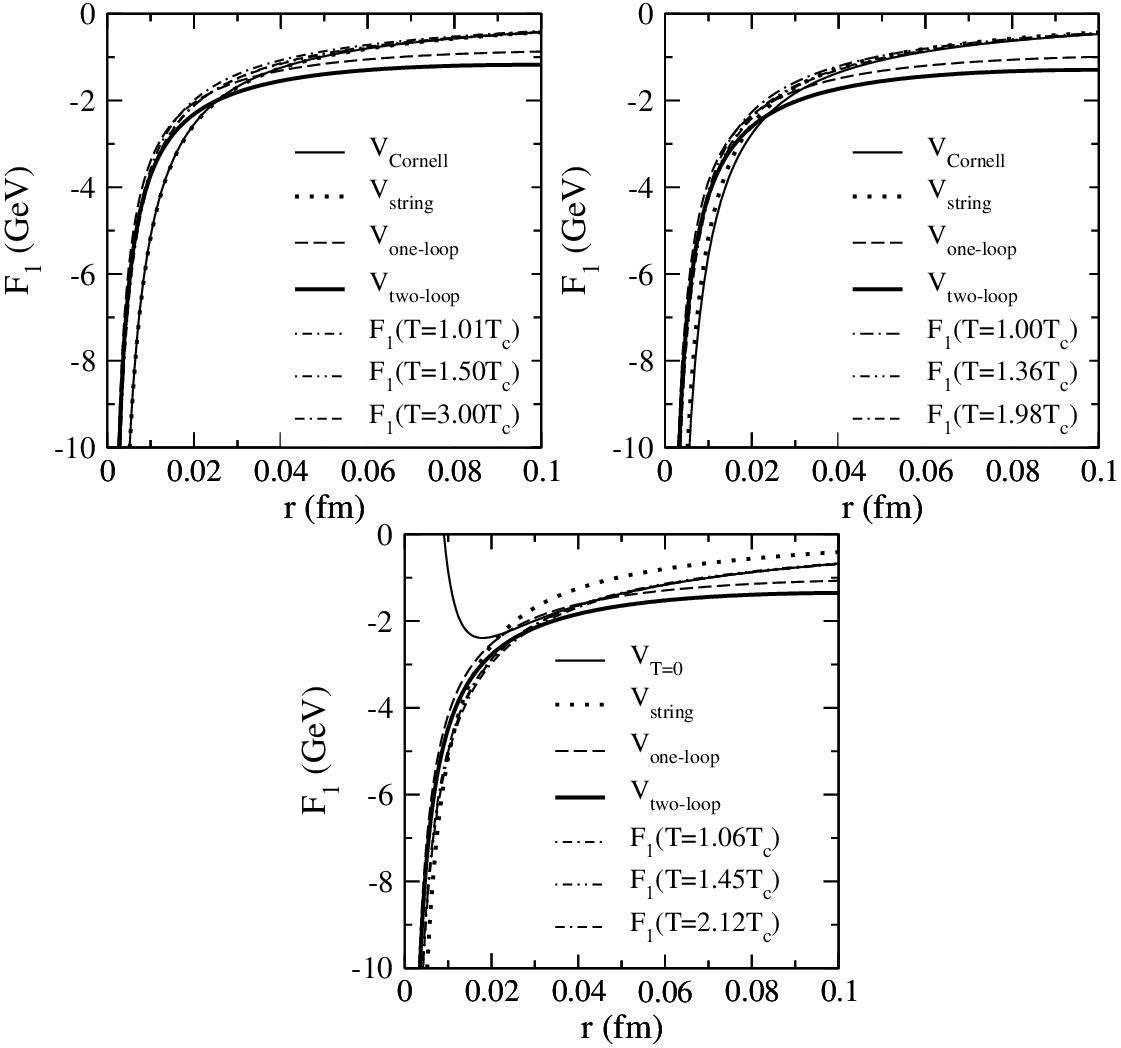}
\caption{The short distance behaviour of the colour singlet free energy
  resulting from our fits for different values of $T/T_c$. The dependence on
  the temperature turns out to be very small at such short distances. We also
  display for comparison the $T=0$ potentials given by different schemes: the
  potentials employed to normalize the data at short distance 
  \protect\cite{Kac4,Petr}, the potential obtained in the Nambu string model
  and the one and two-loop short distance perturbative potentials
  \protect\cite{pet,mel,sch}. }
\label{fig:short} 
\end{center}
\end{figure}

The behaviour of our parameterization of the free energy for $r<0.1$ fm is
displayed, at three different temperatures, in Fig.~\ref{fig:short} where we 
indeed see that spurious short distance thermal effects appears negligible. It
is gratifying (and somewhat surprising) that our curves seems to interpolate
smoothly between the $T=0$ Cornell potential and the short distance
perturbative potential. The Cornell curve reported in the figure is the one
employed in Ref.~\cite{Kac4} to fix the normalization of the free energy at the
different temperatures. As already discussed at the beginning of this section
it was obtained by fitting zero temperature lattice data which cover distances
$r\gtrsim0.05$ fm. Of course for shorter distances the Cornell parameterization
cannot account for running coupling effects (asymptotic freedom) and the
perturbative calculation \cite{pet,mel,sch} should provide more reliable
results. More details on the one- and two-loop perturbative potential used
in Fig.~\ref{fig:short} are given in Appendix~\ref{app:short}.

The curves $V_{\textrm{string}}$ reported in Fig.~\ref{fig:short} refer to the
$Q\overline{Q}$ potential 
\begin{equation}
  V(r)=-\frac{\pi}{12}\frac{1}{r}+\sigma r
\end{equation}
obtained in the Nambu string model, the term $1/r$ arising from the quantum
fluctuations of the flux tube in the transverse directions.

\begin{figure}[t]
\begin{center}
\includegraphics[clip,width=0.9\textwidth]{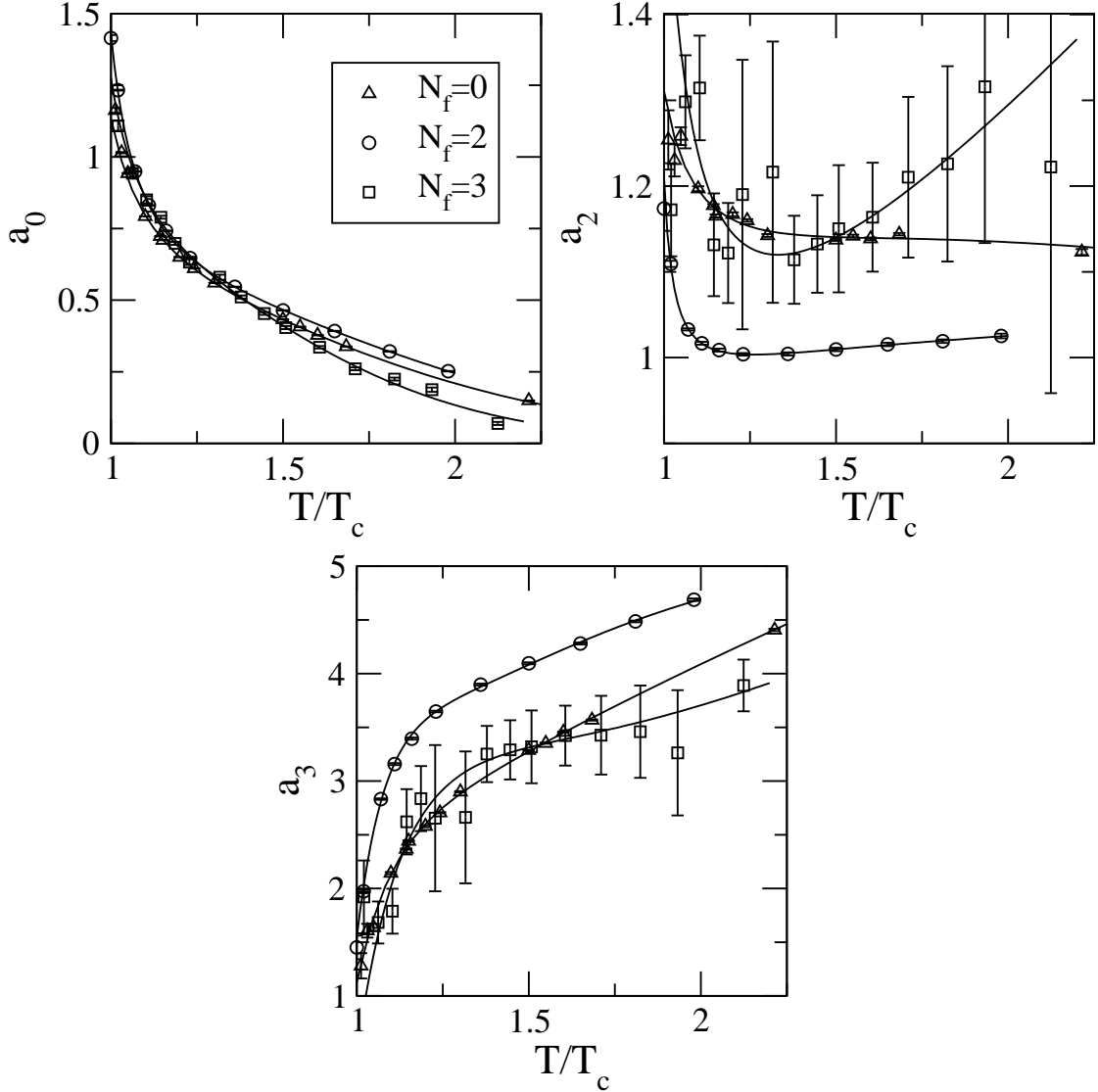}
\caption{Temperature dependence of the parameters $a_0$, $a_2$ and $a_3$
  employed in fitting the color singlet free energy with
  Eq.~(\protect\ref{eq:fittingfunc}). The solid lines are interpolating
  functions. }
\label{fig:tpar} 
\end{center}
\end{figure}

As a next step we have provided a parameterization of the $T$-dependence of
$a_0$, $a_2$ and $a_3$, which allows one to perform analytically the
derivative of $F_1$ with respect to the temperature (actually, for the
calculation of the $Q\bar{Q}$ binding energies the knowledge of $a_0$ is
unnecessary: we show also this contribution to the potential for completeness).
Lacking compelling physical hints to the form of the $T$-dependence,
we have sought for the simplest and smoothest expressions yielding a
satisfactory interpolation of the values calculated from the fit to the lattice
data (see Appendix~\ref{app:Tfit} for details).
The parameters and their interpolations are displayed in Fig.~\ref{fig:tpar}. 

The value of $a_0(T)$ at each temperature is essentially fixed by the
normalization of the data, which has been determined using similar
procedures in all the lattice calculations. Notably this parameter, once
plotted as a function of $T/T_c$, turns out to be fairly close for all values
of $N_f$. On the other hand, the remaining two parameters display some
quantitative flavor dependence, but within the same qualitative pattern. 
An exception to this behavior is represented by $a_2(T)$ for $N_f=3$, although
one should note that the three-flavor case is the one where the
parameterization (\ref{eq:fittingfunc}) has the largest
uncertainties. Moreover, the variation of $a_2(T)$ with $T$ is actually
magnified by the scale of the figure, being generally around 20\% 
for the range of temperatures considered here. As we shall see the resulting
effective potential is rather robust with respect to these variations.

Note also that the strongest variation of the parameters with the temperature
is confined around $T$ very close to $T_c$. Since extracting the internal
energy from $F_1$ involves a derivative with respect to $T$, this is the
temperature domain where the sensitivity to the details of the parameterization
might be high. For this reason we felt it safer to use the resulting
potentials at temperatures larger than $T_c$, say for $T\agt1.05 T_c$, where
they turn out to be stable with respect to changes in the parameterization.

Finally, we recall that the quenched QCD data \cite{Kac1} are actually
available also at temperatures much larger than those for the unquenched cases:
here we limit our analysis up to about twice the critical temperature, since
this is the range where lattice calculations with two and three flavors are
available and, moreover, since this is the range of interest for the problem of
the $J/\psi$ dissociation. Here we just remark that an analysis of the $N_f=0$
data at larger temperatures (till $T\sim 5 T_c$) with the parameterization of 
Eq.~(\ref{eq:fittingfunc}) shows that the fitting parameters maintain, also at
these high temperatures, the trend displayed in Fig.~\ref{fig:tpar}.

\section{$Q\bar{Q}$ bound states in quark-gluon plasma}
\label{sec:qqbar}

The $Q\bar{Q}$ free energy obtained in lattice calculations has been used in
the past as an input for the $Q\bar{Q}$ potential energy in the
non-relativistic Schr\"odinger equation \cite{diga1,diga2,wong1,wong2}. More
recently \cite{Kac1,wong3,shury} it has been recognized that the $Q\bar{Q}$
potential energy can be more appropriately identified with the $Q\bar{Q}$
internal energy (see Eq.~(\ref{eq:pot}))\footnote{Actually, the author of 
  Ref.~\cite{wong3} tries  to disentangle, in the total internal energy
  $U_1(r,T)$, the gluon and the $Q\overline{Q}$ contributions. This leads to a 
  lower value for the $J/\psi$ dissociation temperature with respect to the 
  result $T_d\sim 2T_c$ obtained with the full internal energy. }.

Once the temperature dependence of the color-singlet free energy $F_1$ has been
parameterized, the corresponding color-singlet internal energy $U_1$ is easily
obtained and one can define an effective potential 
\begin{equation}
  V_1(r,T) = U_1(r,T) - U_1(r\to\infty,T)
\end{equation}
to be used in the Schr\"odinger equation
\begin{equation}
  \left[-\frac{\bm{\nabla}^2}{2\mu} + V_1(r,T) \right]\psi(\bm{r},T) =
    \varepsilon(T) \psi(\bm{r},T),\label{eq:scrodeq}
\end{equation}
where $\mu$ is the reduced mass of the $Q\bar{Q}$ system.

\begin{figure}[t]
\begin{center}
\includegraphics[clip,width=0.9\textwidth]{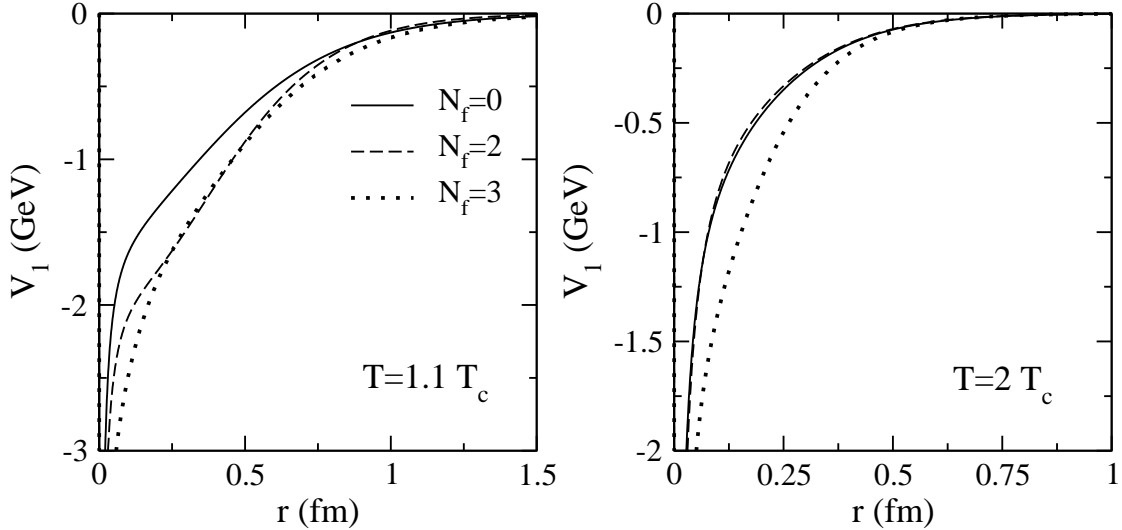}
\caption{ The effective $Q\bar{Q}$ potentials extracted from the unquenched and
  quenched lattice calculations at two different temperatures. }
\label{fig:V1} 
\end{center}
\end{figure}

In Fig.~\ref{fig:V1} we display, at two different temperatures, the effective
$Q\bar{Q}$ potentials that we have obtained by using the parameterizations
discussed in the previous section. At high temperature the form of $V_1$ is the
one typical of a screened Coulomb potential, with the $N_f=3$ potential
providing a stronger attraction; at temperatures close to $T_c$ the shape of
$V_1$ appears somewhat distorted because of the strong temperature dependence
of $a_2$ and $a_3$ (see Fig.~\ref{fig:tpar}), the $N_f=3$ potential being still
more attractive.
Remarkably, the behaviour of the color singlet potential energies obtained with
our procedure appears qualitatively in agreement with the one given in 
Refs.~\cite{Kac2,Kac5}, where the internal energy has been directly calculated
on the lattice for $N_f=0$ and $N_f=2$, respectively.
  
\begin{figure}[t]
\begin{center}
\includegraphics[clip,width=0.8\textwidth]{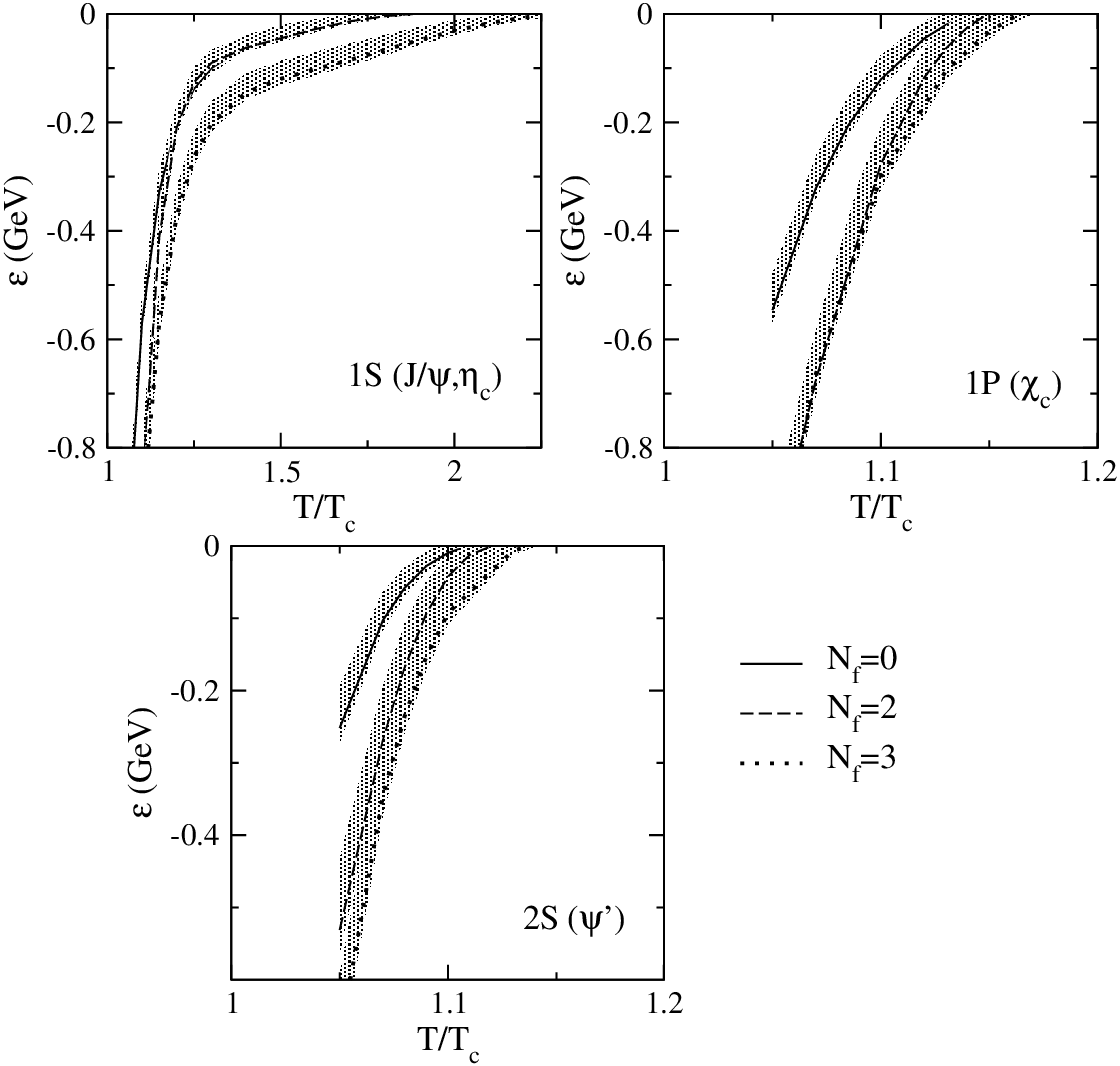}
\caption{ Binding energy of charmonium states above the deconfinement
  temperature. The lines show the results for $m_c=1.3$~GeV; the grey areas 
  around the lines display the variations for 1.15~GeV$<m_c<$1.35~GeV. }
\label{fig:Jpsi} 
\end{center}
\end{figure}

We now employ the effective potential previously derived for the study of the
charmonium and bottomonium spectroscopy above the critical temperature. 

\begin{figure}[p]
\begin{center}
\includegraphics[clip,width=0.75\textwidth]{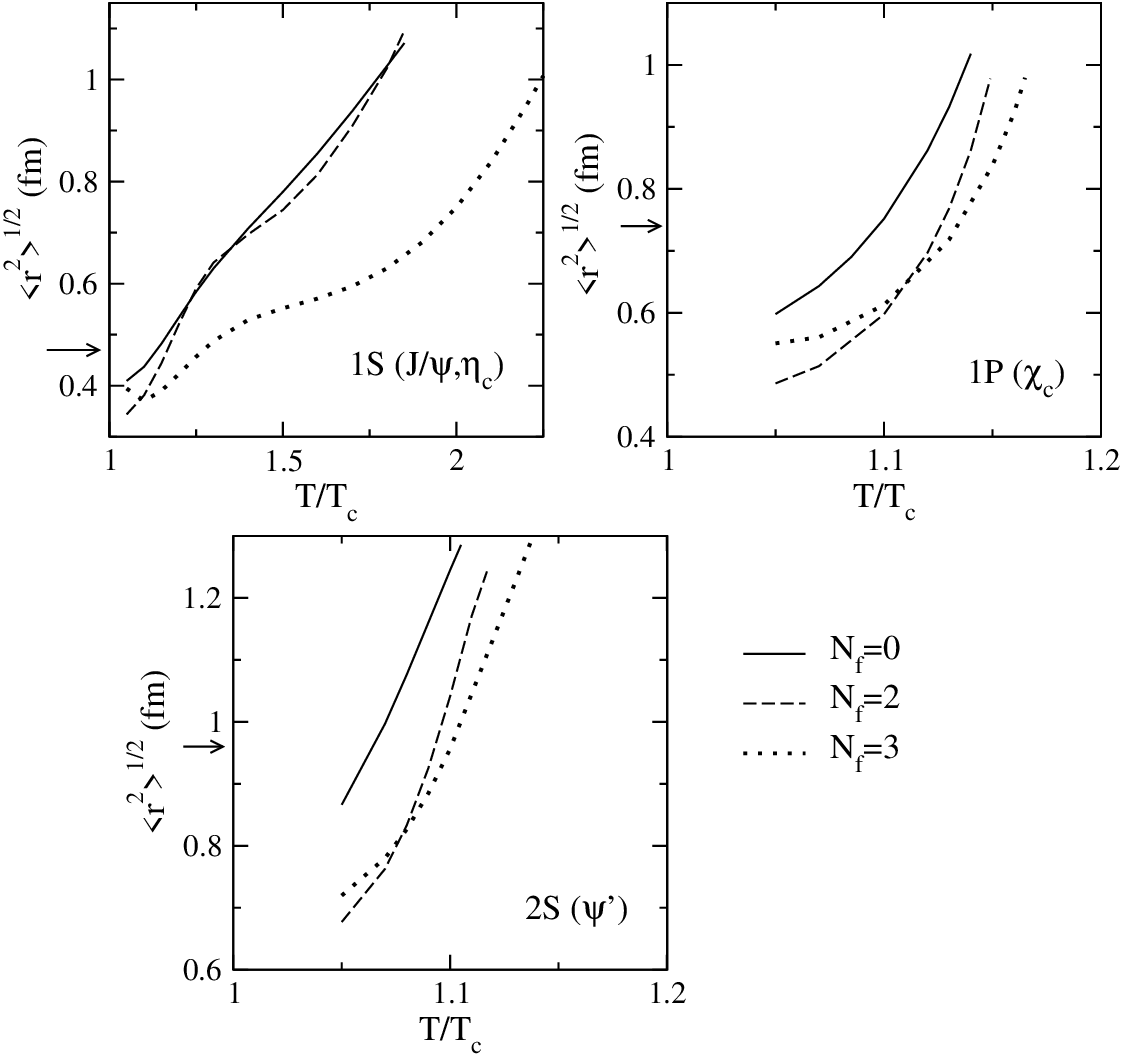}
\caption{Mean square radii of the charmonium states above the deconfinement 
  temperature. The arrows point to the $T=0$ results of
  Ref.\protect\cite{Eic80}. The curves stop where the system is no longer
  bound. } 
\label{fig:Jpsirm}
\vskip 1cm
\includegraphics[clip,width=0.38\textwidth]{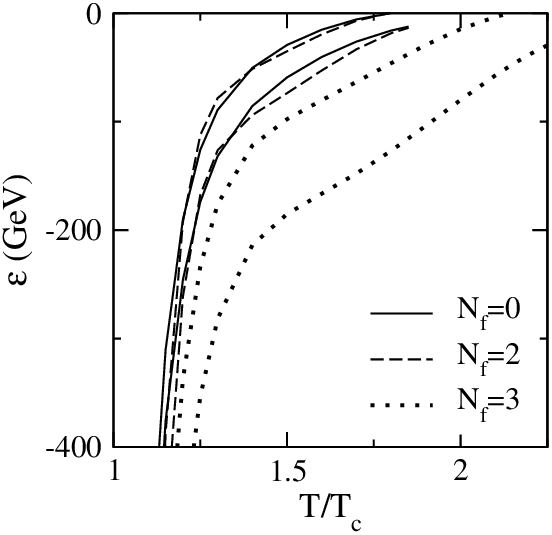}
\caption{ Binding energy of the $1S, J=1$ ($J/\psi$, upper lines) and $1S, J=0$
  ($\eta_c$, lower lines) states above the deconfinement temperature; 
  $m_c=1.3$~GeV. }
\label{fig:Jpsieta} 
\end{center}
\end{figure}

For what concerns the charmonium states we plot the values of the binding
energies (Fig.~\ref{fig:Jpsi}) and of the mean square radii 
(Fig.~\ref{fig:Jpsirm}) of the different bound state solutions of 
Eq.~(\ref{eq:scrodeq}) as functions of $T/T_c$.
For the sake of comparison the results obtained for a different number of
dynamical fermions ($N_f=0,2,3$) are plotted in the same panel.

In our analysis we have chosen $m_c=1.3$~GeV for the charm quark mass; however,
we have also studied the sensitivity of our results to variations of $m_c$ in
the range given in the Particle Data Group (PDG) listings \cite{PDG04}, namely
1.15~GeV$<m_c<$1.35~GeV. 
For reference, in Fig.~\ref{fig:Jpsirm} the arrows point to the $T=0$ mean
square radii obtained in a potential model calculation \cite{Eic80}. 
Note that potential models at $T=0$ typically follow a different philosophy:
they fix the quark mass and the confining potential parameters to reproduce the
mass of the lowest states, whereas here the potential is provided by lattice
calculations and the quark mass is the running mass in the $\overline{MS}$
scheme. 

The $1P$ and $2S$ states turn out to melt at temperatures $1.1T_c\lesssim T_d
\lesssim 1.15T_c$. On the contrary the $J/\psi$ stays bound up to temperatures
$1.7T_c\lesssim T_d\lesssim 2.3T_c$, the precise limits depending upon $N_f$.
The lower bound ($T_d\sim1.7 T_c$) refers to the quenched case and appears in
striking agreement with the limiting value obtained in Ref.~\cite{asa2} through
the study of the $J/\psi$ and $\eta_c$ spectral functions.

Note that the free energy measured on the lattice provides a spin averaged
result of the singlet and triplet channels: hence, in Figs.~\ref{fig:Jpsi} and
\ref{fig:Jpsirm} the $J/\psi$ and $\eta_c$ mesons appear degenerate. 
This degeneracy is expected to be removed by a short range spin-spin force,
whose effect, at $T=0$, is often treated perturbatively assuming a contact
interaction \cite{Bra04}:
\begin{equation}
  H_{ss} = \frac{8\pi}{9} \frac{\alpha_s(\mu)}{m_q m_{\bar{q}}}
    \bm{\sigma}_{q}\cdot\bm{\sigma}_{\bar{q}} \delta(\bm{r}).
\end{equation}
Again, the short range nature of this force makes plausible the assumption that
it is not affected by thermal effects and that it can be employed also at
finite temperatures. Then, for the $J/\psi$ and $\eta_c$ energy shifts one
gets:
\begin{subequations}
\begin{eqnarray}
  \Delta E_{J/\psi} &=& \frac{8\pi}{9} \frac{\alpha_s(\mu)}{m_c m_{\bar{c}}} 
    |\psi(0,T)|^2 \\
  \Delta E_{\eta_c} &=& -\frac{24\pi}{9} \frac{\alpha_s(\mu)}{m_c m_{\bar{c}}}
     |\psi(0,T)|^2,
\end{eqnarray}
\end{subequations}
where for $\alpha_s(\mu)$ we have used the two-loop expression of
Eq.~(\ref{eq:alpha2}), evaluated at $\mu=m_c$. As one can see, the temperature
dependance of this term stems entirely from the value of the $c\bar{c}$ wave
function at the origin.

\begin{figure}[p]
\begin{center}
\includegraphics[clip,width=0.8\textwidth]{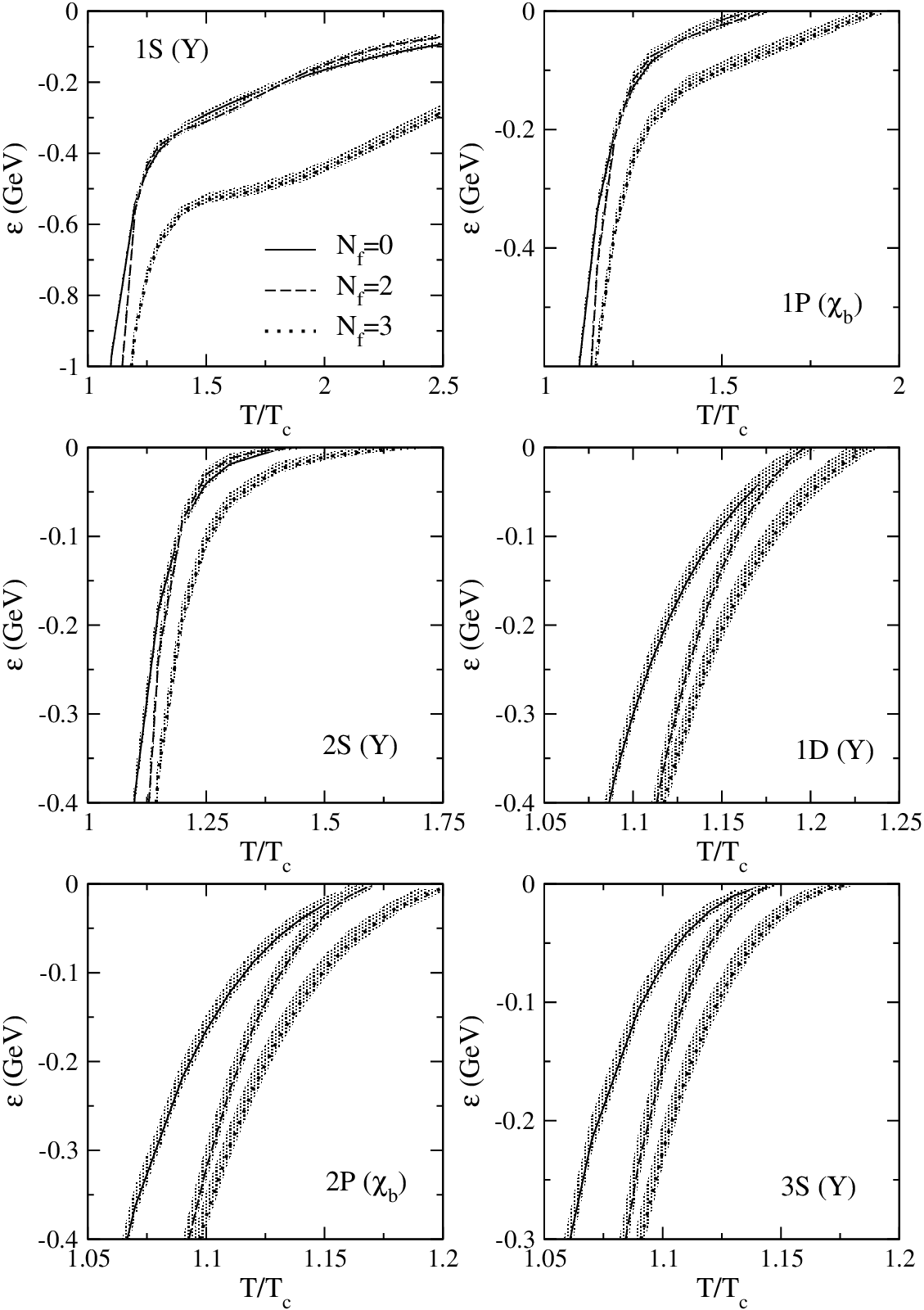}
\caption{Binding energy of the bottomonium states above the deconfinement
  temperature. The lines show the results for $m_b=4.3$~GeV; the grey areas 
  around the lines display the variations for 4.1~GeV$<m_b<$4.4~GeV. } 
\label{fig:Ups} 
\end{center}
\end{figure}
 
\begin{figure}[p]
\begin{center}
\includegraphics[clip,width=0.8\textwidth]{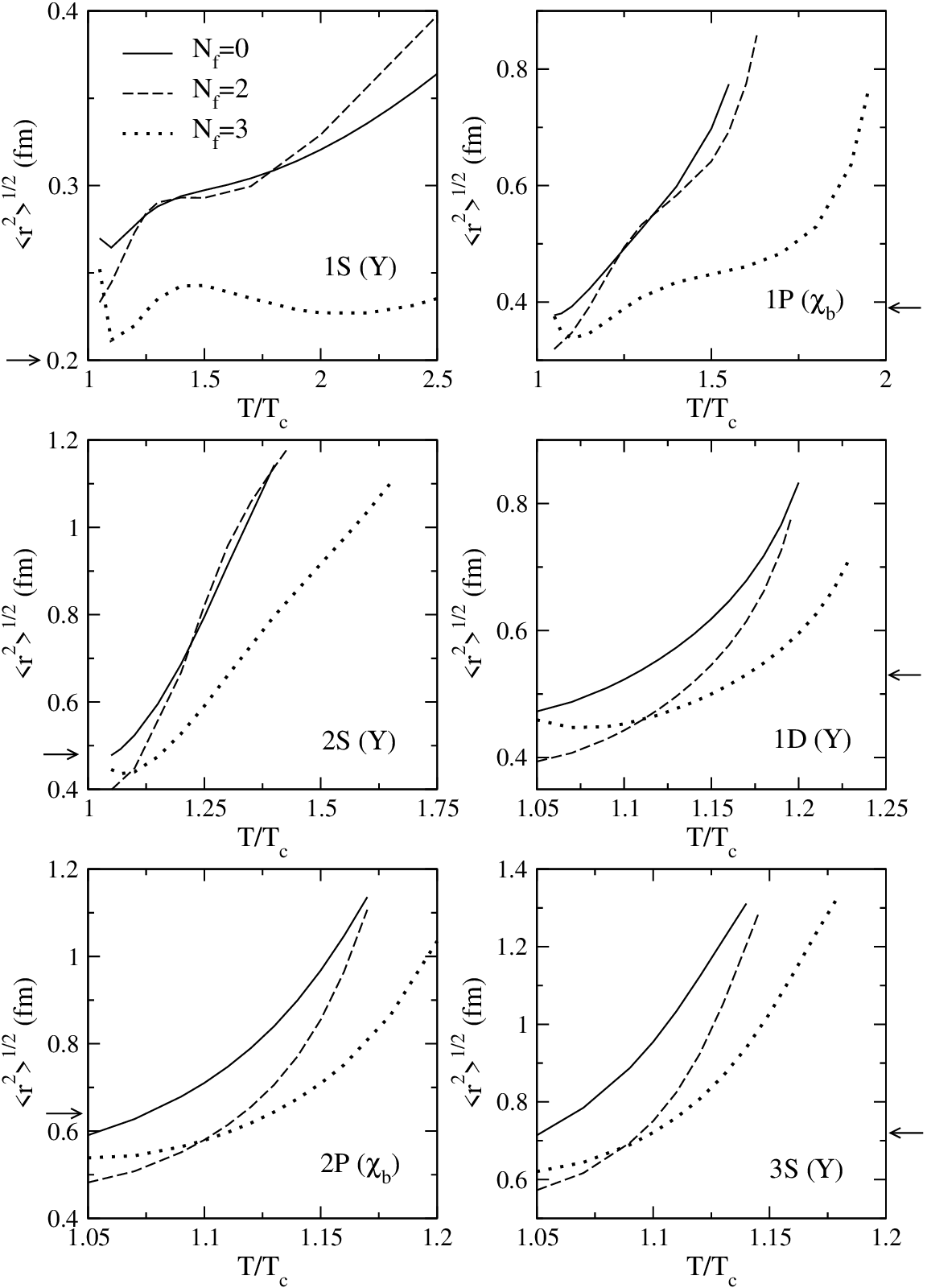}
\caption{Mean square radii of the bottomonium states above the deconfinement 
  temperature. The arrows point to the $T=0$ results of
  Ref.\protect\cite{Eic80}. The curves stop where the system is no longer
  bound. }
\label{fig:Upsrm} 
\end{center}
\end{figure}

We display in Fig.~\ref{fig:Jpsieta} the binding energies of the $J/\psi$ and
$\eta_c$ mesons above the deconfinement temperature. The spin-spin contribution
gives rise to a mild reduction of the dissociation temperature, without
altering the qualitative features of the results. For instance, for $N_f=2$ at
$T=1.05 T_c$ one gets a $J/\psi-\eta_c$ splitting of 145~MeV (the experimental
value at $T=0$ is about 117~MeV), whereas at the $J/\psi$ melting temperature,
$T\cong1.8T_c$, one gets a splitting of 13~MeV (although, strictly speaking,
in this case the perturbative expansion is no longer justified).

In Figs.~\ref{fig:Ups} and \ref{fig:Upsrm} we plot the binding energies and the
mean square radii of the bottomonium bound states above $T_c$. We have chosen
$m_b=4.3$~GeV for the bottom quark mass, but again we have studied the
sensitivity to variations of $m_b$ in the range of the PDG listings,
4.1~GeV$<m_b<$4.4~GeV. The arrows in Fig.~\ref{fig:Upsrm} point to the results
of $T=0$ potential model calculations \cite{Eic80}. Of course, for heavier
quark masses more states than in the charmonium case survive above the
deconfinement temperature.

Both for the charmonium and for the bottomonium states the binding energy gets
smaller and the mean square radius gets larger as the temperature grows. 
This should reflect into a huge increase of the elastic cross section
$\sigma(Q\bar{Q}\rightarrow Q\bar{Q})$ at low energy. In fact, for a small
relative momentum $k$ of the two particles, when in the partial wave expansion
the contribution of the $s$-wave is dominant, the presence of a $l=0$ state
near zero binding energy leads to a dramatic increase of the cross section as
$k\to 0$, according to the formula: 
\begin{equation}
  \sigma_{l=0}\underset{k\to 0}{\sim}\frac{4\pi}{k^2+2\mu|\epsilon|},
\end{equation}
where $\mu$ is reduced mass of the system and $\epsilon$ the energy of the
state near zero binding, no matter whether it is positive (virtual level) or 
negative (bound state). If something analogous happened also for
the heavy-light states (for which lattice data are not available yet) this
would clearly accelerate the thermalization of the heavy quarks. 

From Figs.~\ref{fig:V1},~\ref{fig:Jpsi} and \ref{fig:Ups} the effective
potential obtained with $N_f=3$ appears more attractive. Indeed the critical
temperature $T_c$ in the three cases $N_f=0,2,3$ assumes different values,
hence a naive comparison might not be so meaningful. Furthermore the finite
temperature $N_f=3$ lattice data available so far are affected by larger
errors, so that our parameterization presents larger uncertainties with respect
to the other two cases. 
 
\section{Conclusions}
\label{sec:concl}

The picture of the deconfined phase of QCD for values of the temperature
slightly exceeding $T_c$ --- that is the ones accessible in the heavy-ion
experiments presently performed at RHIC --- has substantially evolved in the
last few years. 

For large values of the temperature ($T\gtrsim3T_c$) a description of the QGP
in terms of a gas of weakly interacting quasiparticles \cite{bla} appears
reliable (even in a regime where the coupling $g$ is not so weak) and supported
by the lattice data for the QGP thermodynamics \cite{thermlat}. 

On the other hand, for temperatures up to $T\sim2T_c$ (namely the ones
currently accessible in the experiments), the matter resulting from the
heavy-ion collisions is nowadays often described \cite{shury} as a
strongly interacting QGP (sQGP). In particular, a striking feature of the QGP
matter obtained at RHIC is its hydrodynamical behaviour \cite{teaney,kolb}
(characterized by a very low viscosity), which manifests itself in particular
in the elliptic flow \cite{olli} observed in non-central collisions: the plasma
obtained at RHIC seems to behave as a nearly ideal fluid whose expansion is
driven by pressure gradients. 

In order to explain the very small mean free path of the plasma particles
required by the hydrodynamical scenario, a picture of the matter obtained at
RHIC in terms of a system of hundreds of loosely bound states of
\emph{quasiparticles} ($q\bar{q}$, $qg$, $gg$...) has been recently proposed
\cite{shury}. In such a framework one has to resort to some assumptions. 
Gluons and light quarks are treated as quasiparticles endowed with quite heavy
thermal masses obtained from lattice calculations and the potential felt by
them in the different color channels is got from the lattice $Q\overline{Q}$
free energy under the hypothesis of \emph{Casimir scaling}. 

The presence of such a pattern of bound states in the range of temperatures
$T_c\lesssim T\lesssim 2T_c$ has been recently questioned on the basis of the
analysis of the correlation between baryon number and strangeness \cite{koch}. 
The evaluation of the correlation coefficient $C_{BS}=-3\langle
BS\rangle/\langle S^2\rangle$ should allow to discriminate between a scenario
in which the relevant degrees of freedom in the QGP are weakly interacting
quark and gluon \emph{quasiparticles} or loosely bound states of the
latters. The lattice data on the off-diagonal quark-number susceptibilities
available so far seem to favour the first hypothesis. 

Here we followed a different approach, starting from the case of a heavy
$Q\overline{Q}$ pair placed in a thermalized QGP and extracting their
interaction from the available lattice calculations. Indeed, we have exploited 
the lattice data for the heavy quark free energies to get information on the
existence of $c\bar{c}$ and $b\bar{b}$ bound states above the deconfinement
transition. We have examined the cases $N_f=0$ \cite{Kac1}, $N_f=2$ \cite{Kac4}
and also $N_f=3$ \cite{Petr}, where lattice data are getting available. For
the color singlet free energy we have adopted a parameterization which accounts
for the effects of asymptotic freedom at short distances and displays an
exponential screening at large distances. From the free energy we have then
extracted the heavy quark potential, to be inserted into the Schr\"odinger
equation. The latter has been solved numerically for the two interesting cases
of charmonium and bottomonium.

For what concerns the charmonium, we have found a dissociation temperature
$T_d\sim 2T_c$ for the 1S states ($\eta_c$ and $J/\psi$). Also the excited
states 1P and 2S appear to melt, but at temperatures slightly exceeding $T_c$. 

On the other hand, the bottomonium spectrum displays a much larger number of
bound states above $T_c$. In particular its ground state turns out to remain
bound in the whole range of temperatures covered by our parameterization. 
By extrapolating the latter at larger $T$ we get a dissociation temperature
$T_d\sim 4\div 6T_c$, depending upon the number of dynamical fermions.  
At fixed $T/T_c$, both for the charmonium and for the bottomonium, the system
turns out to be more bound when the number of light flavors is increased.

Remarkably, in the range of temperatures of experimental interest covered in
this paper a number of loosely bound $Q\overline{Q}$ states exists. 
Since the existence of states near zero binding energy entails a huge increase 
of the elastic cross sections at low relative momenta, this can help the
approach to thermal equilibrium also of the heavy quarks.

An analogous situation occurs in the cross-over from the BCS theory of
superconductivity to the Bose-Einstein condensation, much explored lately in
ensembles of alkaline fermionic atoms confined in a magnetic trap.
Indeed, by smoothly changing the external magnetic field $B$, one induces a
smooth change of the energy of the two interacting atoms: if this energy is
close to the one of a Feshbach resonance, a huge increase in the cross-section
occurs \cite{Tim99,Leg01,Dui04,OHa02}. The striking analogy between $B$ and $T$
as control parameters is self-imposing. 

In connection with the previous discussion on heavy quark thermalization, it 
would be highly desirable to have lattice data also for the $Q\bar{q}$ states, 
which are not yet available.

A phenomenological model has been recently proposed assuming the existence of 
\emph{resonant} (not bound) D- and B- meson states above $T_c$ \cite{van}. 
In this scheme the transverse momentum distributions of the charmed quarks 
(anti-quarks) would approach their thermal equilibrium value much faster, due 
to isotropic resonant scattering on light anti-quarks (quarks). 
Indeed, recent PHENIX results for azimuthally averaged transverse momentum 
spectra of single electrons arising, in Au-Au collisions, from the decay of D-
and B-mesons seem to be compatible with a thermalization scenario \cite{flow}. 

Furthermore, preliminary results from PHENIX and STAR on the elliptic flow of
the above electrons tend to support a picture in which, due to strong
rescattering, charmed quarks reach thermal equilibrium and follow the flow of
the fireball \cite{PHE1,PHE2,STAR1,greco}.
Hence, an extension of the present approach to the $Q\bar{q}$ ($q\bar{Q}$)
states might shed light on the mechanism of thermalization of the heavy quarks.

Coming back to the present results, they offer the relevant possibility of 
evaluating the charmonium (bottomonium) multiplicity produced in heavy ion 
collisions at the experimental conditions of SPS, RHIC and LHC. 

Indeed, a reliable estimate of the dissociation temperature of the different
quarkonia is essential in order to predict how many of them, after being
produced in the hard initial processes, survive in the QGP phase thus
contributing to the final measured yields. The latter of course also contain,
e.~g., the charmonia that might be thermally produced during the hadronization
process by $c\bar{c}$ recombination. 

In particular, the knowledge of the in-medium quarkonium wave function and
binding energy would allow a hopefully reliable estimate of the 
gluon-dissociation cross section of the $J/\psi$ (and hence, via a detailed
balance analysis, also of the cross section for the inverse process). 
Thus a kinetic rate equation accounting for both the dissociation and
recombination processes can be tackled: its solution, depending on how many
$c\bar{c}$ pairs are produced in the initial state, should provide the number 
of $J/\psi$ present at the end of the QGP phase. 
These issues will be addressed in future work. 

\section{Acknowledgments}
We are grateful to O. Kaczmarek and P. Petreczky for providing us their lattice
data. We wish to thank also C.Y. Wong for sending us a revised version of his
paper.

\appendix

\section{Running coupling and perturbative potentials at short distances}
\label{app:short} 

In this appendix we collect the formulas resulting from the perturbative
calculations of Refs.~\cite{pet,mel,sch} of the heavy quark potential
at short distances, where asymptotic freedom guarantees that the perturbative
approach is justified. 

The static QCD potential turns out to be given by:
\begin{equation}
  V(r)=-C_F\frac{\alpha_{\overline{MS}}(\mu)}{r} 
    \left(1+v_1(r,\mu)\frac{\alpha_{\overline{MS}}(\mu)}{\pi}+v_2(r,\mu)
    \frac{\alpha_{\overline{MS}}^2(\mu)}{\pi^2}+\dots\right),
\label{eq:potpert} 
\end{equation}
where \cite{mel} 
\begin{eqnarray}
  v_1(r,\mu)&=& \frac{1}{4} \left[ \frac{31}{9} C_A - \frac{20}{9} T_Fn_f +
  2 \beta_0 \log ( \mu r^\prime) \right], \label{eq:v1r0} \\
  v_2(r,\mu) &=& \frac{1}{16} \left[ \left( \frac{4343}{162} + 4 \pi -
  \frac{\pi^2}{4}
  +\frac{22}{3} \zeta_3 \right) C_A^2 - \left( \frac{1798}{81}+ \frac{56}{3}
  \zeta_3 \right) C_A T_F n_f \right. \nonumber \\ &&
  - \left( \frac{55}{3} - 16 \zeta_3 \right) C_F T_F n_f 
  + \left( \frac{20}{9} T_F n_f \right)^2 
  + \beta_0^2 \left( 4 \log^2 (\mu r^\prime)  + \frac{ \pi^2}{3} \right)
  \nonumber \\
  && \left. + 2 \left( \beta_1 + 2 \beta_0 
  \left( \frac{31}{9} C_A - \frac{20}{9} T_F n_f \right) \right) \log 
(\mu r^\prime) \right]. \label{eq:v2r0}
\end{eqnarray}
In the above $ r^\prime \equiv r \exp ( \gamma_E )$ ($\gamma_E$ is the
Euler-Mascheroni constant), $C_A=3$, $C_F=4/3$, $T_F=1/2$ and
$\zeta_3=\zeta_R(3)$ (the Riemann function of argument 3); moreover, 
$\alpha_{\overline{MS}}(\mu)$ is the QCD running coupling
in the $\overline{MS}$ renormalization scheme coming from the solution of the
RGE: 
\begin{equation}
  \beta(\alpha_s(\mu^2))=\frac{1}{\alpha_s(\mu^2)}
    \frac{\partial\alpha_s(\mu^2)}{\partial\log\mu^2}\equiv-\sum_{n=0}^{\infty}
    \beta_n\left(\frac{\alpha_s(\mu^2)}{4\pi}\right)^{n+1}.
\label{eq:RGE}
\end{equation}
The one-loop perturbative potential is by obtained keeping only the first two
terms of Eq.~(\ref{eq:potpert}) and employing for $\alpha_{\overline{MS}}(\mu)$
the result 
\begin{equation}
\label{eq:alpha2}
  \alpha_{\overline{MS}}(\mu)=
    \frac{4\pi}
    {\beta_0\ln\frac{\displaystyle \mu^2}{\displaystyle
    \Lambda_{\textrm{QCD}}^2}}
    \left(1-\frac{\beta_1}{\beta_0^2}
    \frac{\ln\left(\ln
    \frac{\displaystyle \mu^2}{\displaystyle \Lambda_{\textrm{QCD}}^2}\right)}
    {\ln\frac{\displaystyle \mu^2}{\displaystyle
    \Lambda_{\textrm{QCD}}^2}}\right), 
\end{equation}
which arises from the solution of Eq.~(\ref{eq:RGE}) with the two-loop beta
function (i.e. considering only the first two terms of the series). Indeed the
coefficients $\beta_0$ and $\beta_1$ do not depend on the renormalization
scheme and are given by: 
\begin{equation}
  \beta_0=11-\frac{2}{3}N_f,\qquad\qquad\beta_1=102-\frac{38}{3}N_f.
\end{equation}
On the other hand, for the two-loop perturbative potential one has to keep all 
the terms up to order $\alpha_s^3$ displayed in Eq.~(\ref{eq:potpert}). The
evaluation of the running coupling should include also the three-loop
coefficient $\beta_2$ of the beta-function, which is no more renormalization
scheme independent. In evaluating the two-loop curve in
Fig.~\ref{fig:short} we have used Eq.~(13) of Ref.~\cite{gock}, where the
$\overline{MS}$ scheme is employed.

Clearly, Eq.~(\ref{eq:potpert}) still depends on the parameters $\mu$ and
$\Lambda_{\textrm{QCD}}$. The scale $\mu\sim r^{-1}$ is quite arbitrary: we
chose $\mu=[r\exp(\gamma_E)]^{-1}$ as usually done in the literature. 
For what concerns $\Lambda_{\textrm{QCD}}$ we employed the value suggested by a
recent lattice collaboration~\cite{gock} $\Lambda_{\textrm{QCD}}= 261$ MeV
for all the three cases ($N_f=0,2,3$). 
 
\section{$T$-dependence of the fit parameters}
\label{app:Tfit}

In this appendix we show the functions employed in fitting the temperature 
dependence of the parameters $a_0(T)$, $a_2(T)$ and $a_3(T)$ entering into the
functional form we have adopted for the $Q\bar{Q}$ free energy (see
Sect.~\ref{sec:para}). 

As already mentioned in the text, since we have no phenomenological or
theoretical hints to the functional form of the $T$-dependence of these
parameters, we have tried to use the simplest expressions yielding a smooth fit
and a ``reasonable'' $\chi^2$ (say, of the order of a few units). 

For $a_0(T)$ and $a_2(T)$ we have used the following forms:
\begin{eqnarray}
  a_0 &=& \frac{A^{(0)}_1 x^{A^{(0)}_3} \exp[-A^{(0)}_0 x]}{A^{(0)}_2 x^2 - 1},
    \\ 
  a_2 &=& A^{(2)}_0 \frac{A^{(2)}_1 + A^{(2)}_2 x^2 + x^4}{A^{(2)}_3 + 
    A^{(2)}_4 x + x^3},
\end{eqnarray}
where $x\equiv T/T_c$ and the $A^{(0)}_i$'s and $A^{(2)}_i$'s are fit
parameters. 

Only in the case of $a_3(T)$ we have used an expression inspired by 
perturbative QCD, namely the one for the Debye mass (although, rigorously
this parameter does not represent the Debye mass):
\begin{equation}
  a_3 = \left(1+\frac{N_F}{6}\right)^{1/2} x
    \left( A^{(3)}_0 + \frac{A^{(3)}_1}{x^2} + \frac{A^{(3)}_2}{x^4} +
    \frac{A^{(3)}_3}{x^6} + \frac{A^{(3)}_4}{x^8} \right)
    g_{2-\textrm{loop}}(x), 
\label{eq:a3}
\end{equation}
where $g^2_{2-\textrm{loop}}/4\pi\equiv \alpha_{\overline{MS}}$ is the 2-loop
coupling constant obtained by replacing $\mu/\Lambda_{\textrm{QCD}}$ with
$4.8826 x \cong (2\pi T_c/\Lambda_{\textrm{QCD}}) x$, having used $T_c=202$~MeV
and $\Lambda_{\textrm{QCD}}=261~$~MeV. 
In the case $N_f=3$, because of the larger errors affecting the
parameterization of the free energy, we made the fit of $a_3$ stiffer by
setting $A^{(3)}_4=0$.


\begin{thebibliography}{99}

\bibitem{Na50}    Abreu {\em et al}., 
                  Phys. Lett. B {\bf 477}, 28 (2000).
\bibitem{Na50bis} Na50 Collaboration, 
                  Eur. Phys. J. C {\bf 39}, 335 (2005).
\bibitem{satz}    T. Matsui and H. Satz, 
                  Phys. Lett. B {\bf 178}, 416 (1986).
\bibitem{tew}     R.L. Thews, 
                  hep-ph/0206179,
                  in {\em New States of Matter in Hadronic Interactions},
                  edited by H.T. Elze, E. Ferreira, T. Kodama, J. Rafelski, and
                  R.L. Thews
                  (AIP, New York, 2002).
\bibitem{satz2}   S. Gupta and H. Satz, 
                  Phys. Lett. B {\bf 283}, 439 (1992).
\bibitem{diga1}   S. Digal, D. Petreczy, and H. Satz, 
                  Phys. Lett. B {\bf 514}, 57 (2001).
\bibitem{diga2}   S. Digal, D. Petreczy, and H. Satz, 
                  Phys. Rev. D {\bf 64}, 094015 (2001).
\bibitem{kar}     F. Karsch, 
                  hep-lat/0502014.
\bibitem{Kar97}   D. Kharzeev, C. Lourenco, M. Nardi, and H. Satz,
                  Z. Phys. C {\bf 74}, 307 (1997).
\bibitem{scm}     P. Braun-Munzinger and J. Stachel, 
                  Phys. Lett. B {\bf 490}, 196 (2000). 
\bibitem{gran1}   L. Grandchamp and R. Rapp, 
                  Phys. Lett. B {\bf 523}, 60 (2001).
\bibitem{gran2}   L. Grandchamp and R. Rapp, 
                  Nucl. Phys. A {\bf 709}, 415 (2002).  
\bibitem{dat}     S. Datta, F. Karsch, P. Petreczy, and I. Wetzorke, 
                  Phys. Rev. D {\bf 69}, 094507 (2004).
\bibitem{asa1}    M. Asakawa, T. Hatsuda, and Y. Nakahara, 
                  Nucl. Phys. A {\bf 715}, 863 (2003).
\bibitem{asa2}    M. Asakawa and T. Hatsuda, 
                  Phys. Rev. Lett. {\bf 92}, 012001 (2004).
\bibitem{mc}      L.D. McLerran and B. Svetitsky, 
                  Phys. Rev. D {\bf 24}, 450 (1981).
\bibitem{Kac1}    O. Kaczmarek, F. Karsch, P. Petreczky, and F. Zantow, 
                  Phys. Lett. B {\bf 543}, 41 (2002).
\bibitem{Kac2}    O. Kaczmarek, F. Karsch, P. Petreczky, and F. Zantow,  
                  Nucl. Phys. Proc. Suppl. {\bf 129}, 560 (2004).
\bibitem{Kac3}    O. Kaczmarek, S. Ejiri, F. Karsch, P. Petreczky, and
                  F. Zantow, 
                  Prog. Theor. Phys. Suppl. {\bf 153}, 287 (2004).
\bibitem{Kac4}    O. Kaczmarek and F. Zantow, 
                  hep-lat/0503017.
\bibitem{wong1}   C.Y. Wong, 
                  Phys. Rev. C {\bf 65}, 034902 (2002).
\bibitem{wong2}   C.Y. Wong, 
                  J. Phys. G {\bf 28}, 2349 (2002).
\bibitem{wong3}   C.Y. Wong, 
                  hep-ph/0408020.
\bibitem{shury}   E.V. Shuryak and I. Zahed, 
                  Phys. Rev. D {\bf 70}, 054507 (2004).
\bibitem{raf}     R.L. Thews, M. Schroedter, and J. Rafelski, 
                  Phys. Rev. C {\bf 63}, 054905 (2001).
\bibitem{Petr}    P. Petreczky and K. Petrov, 
                  Phys. Rev. D {\bf 70}, 054503 (2004).
\bibitem{nad}     S. Nadkarni, 
                  Phys.Rev. D {\bf 33}, 3738 (1986).
\bibitem{pet}     M. Peter, 
                  Nucl.Phys. B {\bf 501}, 471 (1997).
\bibitem{mel}     M. Melles, 
                  Phys. Rev. D {\bf 62}, 074019 (2000).
\bibitem{sch}     Y. Schr\"oder, 
                  Phys. Lett. B {\bf 447}, 321 (1999).
\bibitem{Kac5}    O. Kaczmarek and F. Zantow, 
                  hep-lat/0506019.
\bibitem{PDG04}   S. Eidelman {em et al.} (Particle Data Group),
                  Phys. Lett. B {\bf 592}, 1 (2004) (http://pdg.lbl.gov).
\bibitem{Eic80}   E. Eichten, K. Gottfried, T. Kinoshita, K. D. Lane, and 
                  T.M. Yan,
                  Phys. Rev. D {\bf 21}, 203 (1980).
\bibitem{Bra04}   N. Brambilla {\em et al.},
                  hep-ph/0412158.
\bibitem{bla}     J.P. Blaizot, E. Iancu, and A. Rebhan, 
                  hep-ph/0303185,
                  in {\em Quark Gluon Plasma 3},
                  edited by R.C. Hwa and X.N. Wang
                  (World Scientific, Singapore, 2003), p.~60.
\bibitem{thermlat} F. Karsch, E. Laermann, and A. Peikert,
                  Phys. Lett. B {\bf 478}, 447 (2000).  
\bibitem{teaney}  D. Teaney, J. Laueret, and E. Shuryak, 
                  nucl-th/0110037.
\bibitem{kolb}    P.F. Kolb and U.W. Heinz, 
                  nucl-th/0305084,
                  in {\em Quark Gluon Plasma 3},
                  edited by R.C. Hwa and X.N. Wang
                  (World Scientific, Singapore, 2003), p.~634.
\bibitem{olli}    J.Y. Ollitrault, 
                  Phys. Rev. D {\bf 46}, 229 (1992).
\bibitem{koch}    V. Koch, A. Majumder, and J. Randrup, 
                  nucl-th/0505052.
\bibitem{Tim99}   E. Timmermans, P. Tommasini, M. Hussein, and A. Kerman, 
                  Phys. Rep. {\bf 315}, 199 (1999).
\bibitem{Leg01}   A.J. Leggett,
                  Rev. Mod. Phys. {\bf 73}, 307 (2001). 
\bibitem{Dui04}   R.A. Duine and H.T.C. Stoof, 
                  Phys. Rep. {\bf 396}, 115 (2004).
\bibitem{OHa02}   K.M. O Hara, S.L. Hemmer, M.E. Gehm, S.R. Granade, 
                  and J.E. Thomas,
                  Science {\bf 298}, 2179 (2002).
\bibitem{van}     H. van Hees and R. Rapp, 
                  Phys. Rev. C {\bf 71},034907 (2005).
\bibitem{flow}    S. Batsouli, S. Kelly, M. Gyulassy, and J.L. Nagle,
                  Phys. Lett. B {\bf 557}, 26 (2003).
\bibitem{PHE1}    PHENIX Collaboration (S. Kelly {\em et al.}),
                  J. Phys. G {\bf 30}, S1189 (2004).
\bibitem{PHE2}    PHENIX Collaboration (S.S. Adlet {\em et al.}), 
                  nucl-ex/0502009. 
\bibitem{STAR1}   STAR Collaboration (F. Laue {\em et al.}),
                  J. Phys. G {\bf G31}, S27 (2005).
\bibitem{greco}   V. Greco, C.M. Ko, and R. Rapp,
                  Phys. Lett. B {\bf 595}, 202 (2004).
\bibitem{gock}    M. Gockeler {\em et al.}, 
                  hep-ph/0502212.  

\end{thebibliography}
\end{document}